\documentstyle[prl,aps,graphicx,twocolumn]{revtex}

\begin{document}

\title{Phase-coherent frequency measurement of the Ca intercombination line at 657 nm with a
Kerr-lens mode-locked femtosecond laser}
\author{J. Stenger, T. Binnewies, G. Wilpers, F. Riehle and H. R. Telle}
\address{Physikalisch-Technische Bundesanstalt, Bundesallee 100, 38116 Braunschweig, Germany}
\author{J. K. Ranka, R. S. Windeler and A. J. Stentz}
\address{Bell Laboratories, Lucent Technologies, 700 Mountain Ave, Murray Hill, New Jersey 070974, USA}
\date{\today}
\maketitle
\begin{abstract}
The frequency of the Calcium \mbox{$^3$P$_1$---$^1$S$_0$} intercombination line at 657 nm is
phase--coherently
measured in terms of the output of a primary cesium frequency standard using an optical
frequency comb generator comprising a sub--10 fs Kerr-lens mode-locked
Ti:Sapphire laser and an external microstructure fiber for self--phase--modulation. The measured
frequency of \mbox{$\nu_{Ca}$ = 455~986~240~494~276 Hz} agrees within its relative uncertainty of
\mbox{$4 \times 10^{-13}$} with the values previously measured with a conceptually different
harmonic frequency chain and with the value recommended for the realization of the SI unit of length.
\end{abstract}

So far, the use of optical frequency standards for the realization of basic units, e.~g.~of time and
length, was hampered by the difficulties of measuring their high frequencies. As a result, the
most precise frequency values recommended by the Comit\'e International des Poids et Measures
(CIPM) are based on very few frequency measurements \cite{qui99}. Recent progress in
femtosecond pulse generation and the advent of microstructure optical  fibers
\cite{ran00,kni96,bro99}, however, have substantially facilitated such optical frequency
measurements. These schemes \cite{apo00,did00,nie00} start with the highly periodic pulse train
of a Kerr--lens mode--locked laser which corresponds in the to a comb--like frequency
spectrum of equidistant lines. The spectral span of this comb reflects the duration of an
individual pulse while the spacing between the lines is determined by the pulse repetition
frequency. It has been shown, that the fast, spectrally far--reaching Kerr--lens mode--locking
mechanism enforces a tight coupling \cite{ude99} of the optical phases of the individual lines.
As a result, the frequency of any of these lines is given by an integer multiple of the pulse repetition
frequency $f_{rep}$ and a frequency $\nu_{ceo}$ \cite{rei99b,tel99,jon00a} which accounts for the
offset of the entire comb with respect to the frequency origin. If these quantities are known,
the unknown frequency of any external optical signal can be precisely determined by counting
the frequency of the beat--note between this signal and a suitable comb line.
In this Letter, we report the first application of this method to phase--coherent measurements of the
frequency of an optical Ca frequency standard operating
on the $^1$S$_0$---$^3$P$_1$ intercombination transition at 657 nm.

The frequency of
this state--of--the--art standard has already been measured with a conventional harmonic frequency
chain \cite{sch96,rie99}, and the frequency value
of $\nu_{Ca}$ = \mbox{455~986~240~494~150~Hz} has been recommended by the CIPM for the 'Practical
realization of the definition of the metre (1997)'  \cite{qui99} with a relative standard
uncertainty of \mbox{$6 \times 10^{-13}$}.  The present experiments
allow for the first time the comparison of two conceptually different phase--coherent
optical frequency measurements,
the conventional type operating in absolute--frequency domain and the novel type
operating in difference--frequency
domain. This comparison is performed without the need for intermediate transportable standards.

Two of the three required measurements mentioned above, the pulse repetition
rate $f_{rep}$ and the multiple integer factor, are
straightforward to carry out. The measurement of the \underline{c}arrier--\underline{e}nvelope
\underline{o}ffset frequency
$\nu_{ceo}$, which arises from the relative velocity of the carrier phase and
the pulse envelope, is a more demanding task.
Several more or less complex schemes, depending on the comb span available, have
been proposed for the measurement of $\nu_{ceo}$ \cite{tel99}.
The simplest concept requires an octave span of the
frequency comb which is not directly available from our laser. Thus, the available span has to be
expanded, e.~g. by external self--phase--modulation (SPM). Microstructure air--silica fibers, which allow for
tailoring the group velocity dispersion (GVD) properties \cite{ran00} are highly suited for SPM of moderate peak
power pulses available from mode--locked laser oscillators since they provide both lateral and temporal
confinement of the pulses over long interaction lengths. If the octave span is achieved, then
the $\nu_{ceo}$ measurement can be accomplished by second--harmonic generation (SHG)
of the spectral lines in the
low--frequency wing of the comb. Whereas the of the comb lines are shifted by $\nu_{ceo}$
with respect to the origin,
their harmonics are shifted by $2 \nu_{ceo}$. Thus, the beat notes between these
harmonics and the corresponding high--frequency--wing lines of the comb show the desired
component at $\nu_{ceo}$.

\begin{figure}
  \centerline{\includegraphics[width=8.6cm]{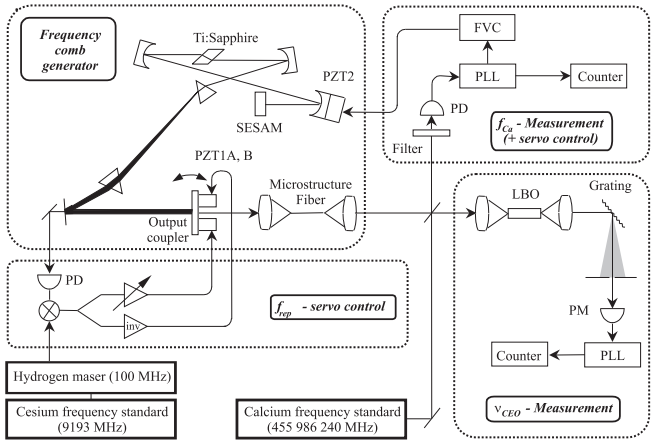}}
  \bigskip
  \caption{Schematic of the experimental set-up.
The frequency comb generator consists of the Ti:Sapphire femtosecond laser and the microstructure fiber.
SESAM denotes the semiconductor saturable absorber mirror, and PZT
piezo transducers. Whereas the repetition rate is phase locked to the
hydrogen maser, $f_{Ca}$ and $\nu_{ceo}$ are counted.
LBO denotes the LiB$_3$O$_5$--frequency doubling crystal, PD photo detectors, PM photomultiplier,
PLL phase--locked loop tracking oscillator, and FVC a frequency--to--voltage converter, respectively.}
\end{figure}

As depicted in \mbox{Fig.~1},
our measurement set--up consists of four parts: the frequency comb generator (i),
detectors and electronics for servo control and/or data acquisition of the pulse repetition frequency (ii),
of the beat note between the laser and Ca--standard radiation (iii) and of the
carrier--envelope--offset frequency (iv).

i) The Kerr-lens mode-locked (KLM) Ti:Sapphire laser is similar to that
described in \cite{sut99}. The laser generates pulses with a duration of  \mbox{$\sim$ 10~fs},
a repetition rate \mbox{$f_{rep}$~=}~\mbox{100~MHz}, an
average power of \mbox{120~mW} and a spectrum centered at \mbox{790~nm}.
The group velocity dispersion (GVD) of its
cavity is controlled by double--chirped mirrors and an intracavity fused silica prism pair.
A similar external
prism pair (not shown in  \mbox{Fig.~1}) is used for pulse recompression.
Approximately 30 mW of the laser output is
coupled into a ~10 cm long piece of microstructure fiber with a core diameter of \mbox{1.7~$\mu$m} and a
zero--GVD wavelength of \mbox{780~nm} \cite{ran00}. The output spectrum of the fiber was measured to cover the
range from \mbox{520~nm} to \mbox{1100~nm}, while the comb structure with distinct lines spaced by
\mbox{100~MHz} is expected to be
preserved even in the extreme wings of the SPM--broadened spectrum. This was confirmed by measurement
of the fringe--visibility (after appropriate optical filtering) at the output of a two beam interferometer with
\mbox{100~MHz} fringe spacing yielding a \mbox{100~\%} visibility throughout the entire comb.\\
ii) The repetition frequency of the laser was detected with a fast \mbox{Si~PIN} photo
diode yielding a \mbox{100~MHz}
signal with a S/N--ratio of \mbox{140~dB} (BW = 1~Hz). It was phase--locked to the \mbox{100~MHz}
output signal of a hydrogen
maser which was controlled by the PTB primary Cs standards. The actuator of this servo loop is a double piezo
device (PZT1A and B) which tilts the output coupler where the spectral components are laterally
displaced by the intra--cavity double prism GVD compensator. Tilting of the output coupler
predominantally modifies the resonator
round--trip group delay \cite{rei99b} which, in turn, determines $f_{rep}$.\\
iii) The frequency  $f_{Ca}$ of the beat note between the radiation from the Ca--standard
(at optical frequency $\nu_{Ca}$) and
the nearest comb mode was both measured and long--term stabilized. For this purpose, the output of the Ca
standard was combined with the emission from the output of the microstructure fiber, which was filtered by an
interference filter. The beat note was detected by a \mbox{Si~PIN} photo diode.
Its output signal with a typical
S/N--ratio of \mbox{30~dB} (\mbox{BW~=}~\mbox{100~kHz})
was filtered by a phase--lock--loop tracking oscillator (PLL) and subsequently
counted by a totalizing counter. The beat signal frequency was kept within the hold--in range of the PLL by an
additional servo loop controlling the length of the Ti:Sapphire--laser resonator with the help of a piezo transducer
(PZT2). The error signal of this slow loop was generated by a frequency--to--voltage--converter (FVC) which
monitored the output  frequency of the PLL.
Unwanted cross--talk between the $f_{rep}$ and $f_{Ca}$ servo loops was minimized by adjustment
of the lever point of the $f_{rep}$--tilting actuator to the lateral position of the red comb components
at \mbox{657~nm}. This was accomplished
by choosing appropriate gain values (of opposite sign) for the piezo transducers PZT1A and B.\\
iv) The frequency $\nu_{ceo}$ was measured, as mentioned above, by external self--phase
modulation in the microstructure fiber and
subsequent second--harmonic generation in a nonlinear optical crystal. The infrared portion of the fiber
output around \mbox{1070~nm} was frequency doubled with a non--critically phase--matched,
\mbox{10~mm} long LiB$_3$O$_5$ (LBO) crystal,
which was heated to about \mbox{140~$^\circ$C} in order to fulfill the \mbox{type-I} phase--matching condition
for this wavelength. The
beat note between the resulting green SHG signal and the green output of the fiber was detected by a photo
multiplier (PM) after spectral and spatial filtering both fields with a single mode fiber and a
\mbox{600~l/mm} grating, respectively.
The PM output signal with a S/N ratio of up to \mbox{40~dB}(\mbox{BW~=}~\mbox{100~kHz})
was filtered with a
second tracking oscillator and counted by a second totalizing counter. Both counters were synchronously gated
by the same \mbox{1~Hz} clock signal derived from the 100 MHz hydrogen maser signal.

The optical Ca frequency standard has been described elsewhere \cite{rie99,rie99a},
and only the relevant details shall be given here. This
standard is operated in a different building and is linked to the Ti:Sapphire laser by a  \mbox{150~m} long
polarization--preserving  single--mode fiber.
Two independent systems are available, producing clouds of between $10^6$ and $10^7$ ballistic Ca atoms
which are released from magneto--optical traps (\mbox{MOT~1} and \mbox{MOT~2}).
Each trap is loaded for \mbox{15~ms}. Then the
trapping lasers  at \mbox{423~nm} and the magnetic quadrupole fields are switched off,
and a homogeneous
magnetic field of \mbox{0.23~mT} is switched on for spectral separation of the Zeeman components.
At the same time the cloud starts to expand and to fall down due to its thermal velocity ($T \cong$~\mbox{3~ mK})
and gravity,
respectively.  After a settling time for the magnetic field of \mbox{0.2~ms} the atoms are interrogated
by pulses from a frequency--stabilized dye laser. For the frequency
measurements \mbox{MOT~1} was employed, in which the low velocity atoms are directly captured from a
thermal beam
by three mutually perpendicular counterpropagating beams with two different frequencies.
The atoms were excited in a time--domain analogue of the optical Ramsey excitation by two pulses separated
by \mbox{216.4~$\mu$s} followed by two pulses with the same separation from a counterpropagating beam
\cite {rie99a}. The
frequency of the interrogating laser field was stabilized to the central fringe (FWHM of the fringes~\mbox{1.16~kHz})
of the interference structure that occurs when the fluorescence of the excited atoms is measured as function of
the laser frequency.
The uncertainty of the frequency of the Ca optical frequency standard during both measurement series was
estimated to \mbox{53~Hz}. A more detailed uncertainty budget is given in \cite{rie99a}.
Several tests have been
performed to check the validity of this estimate. By reversing the temporal order of the excitation pulses a shift
of a few Hertz was determined.
The frequency of the laser stabilized to atoms of \mbox{MOT~1} was compared with its
frequency when stabilized to a second \mbox{MOT~2}.  Here, the atoms of an effusive beam are decelerated in a
Zeeman slower and these slow atoms are deflected and directed towards the magneto--optical trap \cite{kis94}.
The frequency difference measured when the laser was stabilized alternatively to atoms from \mbox{MOT~1} and
\mbox{MOT~2} by three pulses of a standing wave was \mbox{$12\pm 15$~Hz}.
An additional laser field at \mbox{672 nm} was used in \mbox{MOT~1} to repump the atoms which are
lost via the $^1$D$_2$ state \cite{oat99}. The magnetic field
of 0.23 mT caused a second order Zeeman shift of the Ca frequency of \mbox{+3.1~Hz}.

\narrowtext
\begin{table}
 \begin{tabular}{lccc}
      & $\nu_{Ca} -$ & 1$\sigma$ [Hz] & relative 1$\sigma$\\
     & $\nu_{Ca (CIPM)}$ [Hz] &  & uncertainty\\
    \tableline
    Series A & 100 & 328 & $7.2\times 10^{-13}$ \\
    Series B & 148 &132 & $2.9\times 10^{-13}$ \\
    Series A+B & 126 & 180 & $4.0\times 10^{-13}$ \\
    Previous Meas. \cite{rie99a} & -20 & 113 & $2.5\times 10^{-13}$ \\
    CIPM recomm. \cite{qui99} & 0 & 270 & $6.0\times 10^{-13}$ \\
  \end{tabular}
  \caption{The differences of the measured Ca $^3$P$_1$ --- $^1$S$_0$ transition frequencies in respect to the
value recommended by CIPM   and the absolute and relative uncertainties.
Series A and B refer to the
frequency comb generator and previous measurement refers to the harmonic chain.}
\end{table}

Two measurement series were carried out on different days with a total measurement time of 3000 and
\mbox{4200~s}, respectively. On these days the frequency deviation of the hydrogen maser and the Cs clock
resulted in deviations of the measured Ca frequency of \mbox{+26~Hz} and \mbox{24~Hz}, respectively.
The deviations of the corrected measured mean values from the
CIPM--recommended value of $\nu_{Ca (CIPM)}$ = \mbox{455~986~240~494~150~Hz}
are listed in Tab. 1 together with the corresponding relative 1$\sigma$ uncertainties,
which comprise the estimated relative frequency uncertainties of the standards and those of the frequency
measurements.
The values of the previous measurement \cite{rie99} using the harmonic frequency chain
and the CIPM values \cite{qui99} are shown for comparison.

\vspace{1.7cm}
\begin{figure}
  \centerline{\includegraphics[width=9cm]{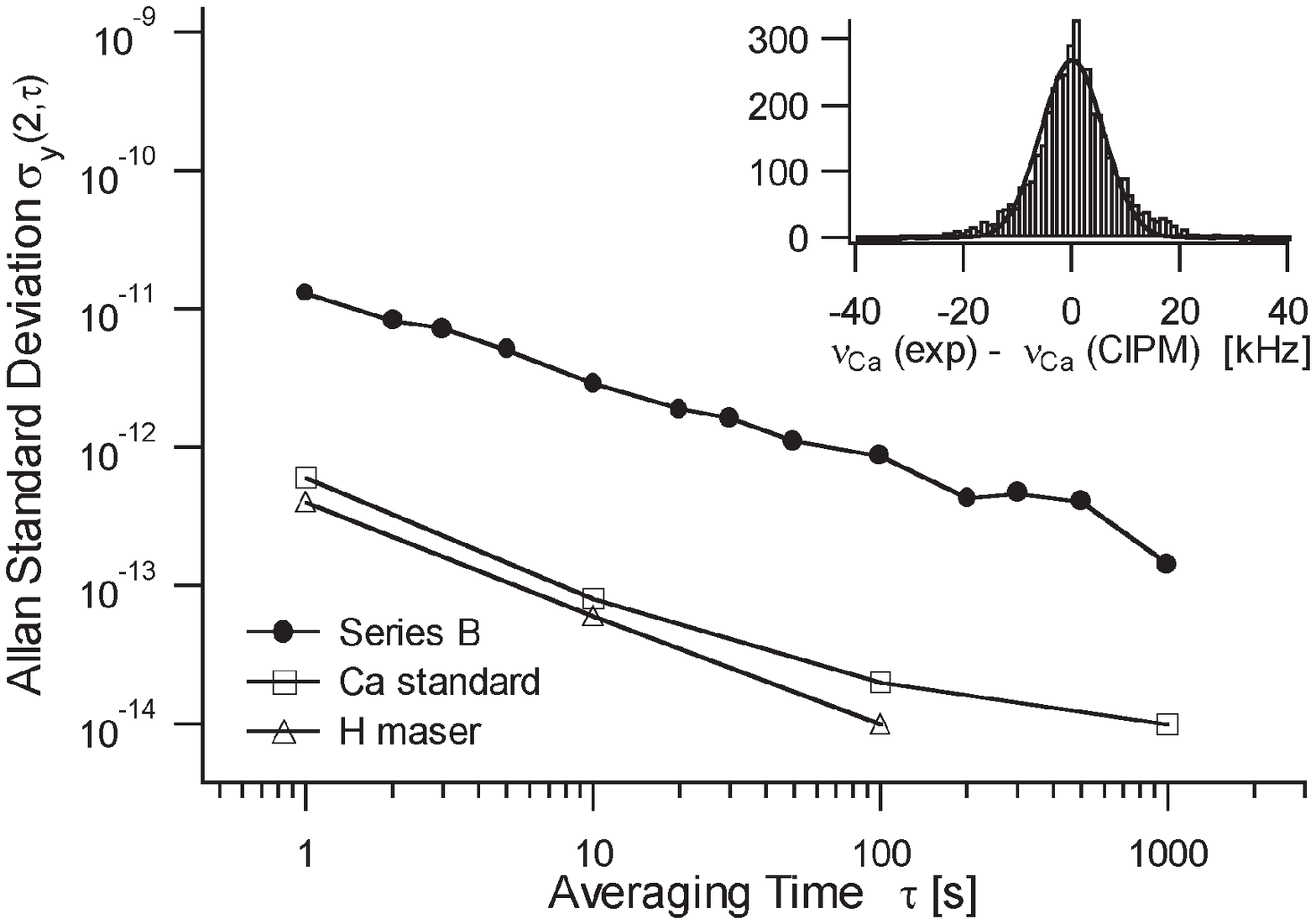}}
  \bigskip
  \caption{The Allan standard deviation of measurement series B is plotted versus
the averaging time $\tau$ together with the
frequency instabilities of the Ca standard and the hydrogen maser. The inset shows the
distribution of the measured values of series B (averaging time \mbox{$=1$~s},
bin-width \mbox{$=1$~kHz}, \mbox{$\sigma=5.75$~kHz}).   }
\end{figure}

\mbox{Fig.~2} shows the Allan standard deviation of the data of series B together with the corresponding frequency
instabilities of the Ca standard \cite{rie99b} and those of the hydrogen maser \cite{bau00}.
The inset of \mbox{Fig.~2} shows a histogram of
the data of series B (\mbox{1~s} averages) with respect  to the CIPM--recommended frequency.
The center frequency  of the fitted Gaussian of $\nu_{Ca (CIPM)}$ + 46 Hz agrees within the uncertainty
with the mean value of series B.
The instability of the frequency measurement process is about one order of magnitude larger than the
frequency instability of the Ca standard and of the hydrogen maser. So far,
the measurements are still limited by the non--optimum detection of the
frequency fluctuations of the Ti:Sapphire laser.
Substantial improvement of the measurement process
can be expected if the instantaneous residual phase error between $f_{rep}$ and the hydrogen maser
reference signal is simultaneously measured, i.e. at a high harmonic of $f_{rep}$.

To conclude, the frequency of an optical frequency standard based on the Calcium
\mbox{$^3$P$_1$---$^1$S$_0$} intercombination transition
at \mbox{657~nm} has been measured in terms of the output of a primary cesium frequency standard using a
novel type of optical frequency chain. Accurate frequency measurements of this optical standard are of
particular importance since it represents one of the realizations of the unit of length with lowest uncertainty.
Furthermore, the optical frequency of the Ca standard is used as a reference for highly accurate frequency
measurements of other optical or ultraviolet transitions \cite{vog00}.
The frequency value presented in this Letter agrees within its relative uncertainty with the CIPM--recommended
value, which is based on a measurement with a harmonic frequency chain. The agreement justifies the confidence
in the data obtained from both measurement schemes. The advantages of the novel scheme employed here,
such as the
substantially reduced complexity and the accurately known, dense grid of reference frequencies throughout the
visible and near infrared range, allows a variety of new applications including the realization of optical clocks or
ultra--precise measurements in fundamental physics.

We gratefully acknowledge financial support from the Deutsche Forschungsgemeinschaft through SFB 407,
experimental assistance by B.~Lipphardt and H.~Schnatz, and help for the set--up of the Ti:Sapphire laser
by G.~Steinmeyer and U.~Keller.

\end{document}